\documentclass[12pt,prl,english,twocolumn,reprint]{revtex4-1}
\usepackage{times}
\usepackage{array}
\usepackage{longtable}
\usepackage{float}
\usepackage{amsmath}
\usepackage{graphicx}
\usepackage{amssymb}
\usepackage{color}
\usepackage[FIGTOPCAP]{subfigure}

\makeatletter
\usepackage{fancyhdr}
\usepackage{babel}
\makeatother
\begin{document}

\title{Runaway electron current reconstitution after a non-axisymmetric magnetohydrodynamic flush}

\author{Christopher J. McDevitt}
\affiliation{Nuclear Engineering Program, University of Florida, Gainesville, FL 32611}
\author{Xian-Zhu Tang}
\affiliation{Theoretical Division, Los Alamos National Laboratory, Los Alamos, NM 87545}

\date{\today}

\begin{abstract}

  Benign termination of mega-ampere (MA) level runaway current has been
  convincingly demonstrated in recent JET and DIII-D experiments,
  establishing it as a leading candidate for runaway mitigation on
  ITER. This comes in the form of a runaway flush by parallel
  streaming loss along stochastic magnetic field lines formed by
  global magnetohydrodynamic instabilities, which are found to
  correlate with a low-Z injection that purges the high-Z impurities
  from a post-thermal-quench plasma. Here we show the competing
  physics that govern the post-flush reconstitution of the runaway
  current in a ITER-like reactor where significantly higher current is
  expected. The trapped ``runaways'' are found to dominate the seeding
  for runaway reconstitution, and the incomplete purge of high-Z
  impurities helps drain the seed but produces a more efficient
  avalanche, two of which compete to produce a 2-3~MA step in current
  drop before runaway reconstitution of the plasma current.

\end{abstract}

\maketitle



The generation and evolution of runaway
electrons (RE) have been extensively studied in a variety of contexts
including atmospheric plasmas~\cite{dwyer2014physics}, solar
flares~\cite{holman1985acceleration,Aschwanden:2002} and magnetic
fusion devices~\cite{Knoepfel:1979}. These highly relativistic
electrons have recently emerged as a topic of particular interest and importance to
the magnetic fusion community. This is due to the possibility that a
large population of REs may be inadvertently generated during a
tokamak disruption~\cite{Hender:2007}. Due to their high energy, often
in excess of $10\;\text{MeV}$, such electrons have the potential to
impart substantial damage to plasma facing
components (PFC)~\cite{Hender:2007}.

A major step toward mitigating the threat posed by REs has recently
been taken. Specifically, recent experiments on
JET~\cite{reux2021demonstration} and DIII-D~\cite{paz2021novel} have
shown that (1) massive injection of deuterium into a
post-thermal-quench plasma can purge the high-Z impurities and
subsequently trigger large scale magnetohydrodynamic (MHD) modes, (2)
independent of the specific MHD modes involved, which are different in
DIII-D and JET experiments, parallel streaming loss in the resulting
stochastic magnetic field lines leads to the expulsion of a preformed
beam of REs, and (3) the spread of the escaping
runaways on the PFCs is sufficiently broad such that no appreciable localized
heat load is observed. A most striking feature of this scheme is its
compatibility with a thermal quench in which plasma energy loss is
dominated by impurity radiation. The high-Z impurities could be
introduced into the plasma accidentally, for example, in the form of a
tungsten flake, or deliberately, for example, through pellet
injection, to mitigate the thermal load on the PFC.  The byproduct of
such strong radiative cooling is a robust Ohmic-to-runaway current
conversion. On ITER, a 15 mega-ampere (MA) plasma current discharge
could produce a post-thermal-quench plasma of over 10~MA RE
current~\cite{iter-basis-chapter3-nf-1999,Hender-etal-nf-2007,Martin:2017,vallhagen2020runaway}.
Safely terminating such a large runaway current has been a
particularly difficult challenge, for which the 3D MHD flush of
REs associated with high-Z impurity purge by massive deuterium
injection offers an attractive solution.

An issue that is anticipated, but has not materialized with
certainty in experiments to date, is the RE current
reconstitution after the spontaneous expulsion of REs by the 3D
magnetic fields.  As long as the flux surfaces reheal after the
self-excited 3D MHD event, the Ohmic plasma current after the runaway
flush is similarly susceptible to Ohmic-to-runaway conversion, just
like the plasma after the initial radiative thermal quench.  One
difference is the reduced impurity content after the purge due to
deuterium injection. If the purge is sufficiently complete and the
remnant deuterium density is not too high, Ohmic heating can offset
the radiative and transport losses, and reheats the plasma so the
parallel electric field $E_{\parallel\eta}=\eta j_\Vert$ can drop
below the runaway avalanche threshold $E_{AV}.$ If this could be
maintained over the rest of the current quench, effective runaway
``avoidance'' would have been achieved.  The primary challenge in that
scenario becomes a goldilock requirement on ion densities such that
radiative cooling and Ohmic heating would offset each other to lock
the plasma to a temperature~\cite{mcdevitt-tang-arXiv-2022} that is
consistent with an Ohmic current decay time in the range of 50-150~ms
for ITER.~\cite{lehnen2015disruptions,Hollmann-etal-PoP-2015}

In the more conservative and perhaps more likely scenario that the
impurity purge is inadequately complete and electron reheating is
insufficient to reach $E_{\parallel\eta} < E_{AV},$ the same avalanche
physics can drive runaway current reconstitution.  The key question
becomes how much the plasma current would drop before another
Ohmic-to-runaway current conversion is completed. The answer to this
question would dictate the issues one must face in the post-flush
mitigation designs.

This Letter lays out the fundamental physics considerations underlying
the answer to the question of runaway reconstitution after an MHD
flush, which are of critical importance to a tokamak reactor like
ITER.  The interesting finding is that the runaway current reconstitution follows
the same avalanche growth physics as the initial runaway current
formation, but the runaway seeding takes place via a new route that makes the
runaway reconstitution a far more robust process than the initial
runaway plateau formation immediately following the plasma thermal
quench. The new feature is the ``trapped runaway'' population in the
runaway plateau phase that is greatly enhanced by the high-Z
impurities before their purge by massive deuterium injection. In a
mitigated post-TQ plasma, the electron temperature is radiatively
clamped to a very low temperature, possibly in the few eV range.  It
is due to the partial screening effect that weakly ionized
high-atomic-number impurities can efficiently scatter the (passing)
runaways into the trapped region.  The radial loss of trapped
electrons thus formed, in sharp contrast to that of passing runaways,
are insensitive to the stochastic magnetic fields in an MHD event.
Furthermore, because of their high energies, the trapped runaways are
resilient against both collisional slowing down and collisional
detrapping during the transition period from the purge to the eventual
flux surface rehealing. This plants the seed for a robust runaway
reconstitution of the plasma current once the flux surfaces are
rehealed after the 3D MHD event.

In a post-flush plasma with $E_{\parallel\eta} > E_{AV},$ the runaway current
reconstitution follows the same avalanche physics as during the initial formation of the RE plateau, i.e.
\begin{equation}
I^{(max)}_{RE} = 2\pi \int^a_0 dr r j^{\left( seed \right)}_{RE}
\left( r \right) 10^{\left| \psi \left( r \right) / \psi_{10} \left(
  r\right) \right|} , \label{eq:0}
\end{equation}
which we have written for a simple geometry with circular flux
surfaces, $r$ is the radial variable, and $a$ is the minor radius. The
amount of poloidal flux required for an order of magnitude increase in
runaway population is labeled as $\psi_{10},$ itself a function of
minor radius $r.$ Eq.~(\ref{eq:0}) expresses the amount of RE
current that could be generated if all of the available poloidal flux
were used to amplify the RE seed, and thus
corresponds to an upper bound on the amount of RE current that can be
generated for a given RE seed.
To minimize runaway current reconstitution, one aims for a higher flux
consumption rate $\psi_{10}(r)$ and a smaller runaway seed
$j_{RE}^{(seed)}.$ A post-flush plasma appears to be favorable on both
accounts: (1) reduced high-Z impurity density due to its purge by
hydrogen injection decreases $\psi_{10};$ (2) a lower $T_e$ implies
ineffectiveness of both the Dreicer flux and hot tail formation in
seeding the runaways, which points to the most optimistic scenario in
which tritium decay and Compton scattering set the minimal runaway
seeding in the nuclear phase of ITER. The reality tilts toward the
opposite, and the spoiler turns out to be magnetic trapping of high
energy electrons.

\begin{figure}
\begin{centering}
\includegraphics[scale=0.5]{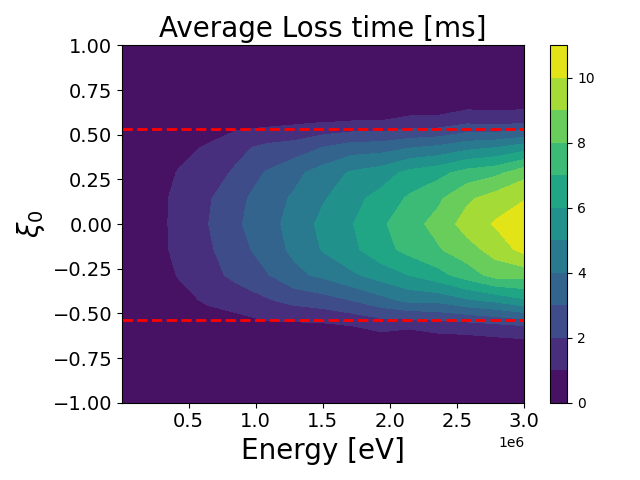}
\par\end{centering}
\caption{Average loss time of electrons in an imposed 3D magnetic
  field. The dashed red lines indicate the location of the
  trapped-passing boundary. The imposed perturbation has the form
  $\delta \mathbf{B} = \nabla \times \left( \alpha
  \mathbf{B}_{eq}\right)$, where $\mathbf{B}_{eq}$ is the equilibrium
  magnetic field, $\alpha = \sum_{m,n} \alpha_{m,n} \left( r \right)
  \cos \left( m\theta - n \varphi + \delta_{m,n}\right)$, with $n=
  \left[ 1, 6 \right]$, $m = \left[ 1, 24 \right]$, and the magnitude
  of $\alpha_{m,n} \left( r \right)$ was increased until a magnetic
  field without any detectable integrable regions was achieved. The
  electrons were all initialized at $r/a=0.5$. The plasma parameters
  were taken to be $n_D=5\times 10^{20}\;\text{m}^{-3}$, $E/E_c = 50$, an ITER like
  plasma with $B_0 = 5.3\;\text{T}$ and minor radius of
  $a=200\;\text{cm}$ was assumed.}
\label{fig:I1}
\end{figure}

{\it Magnetic trapping in a stochastic field.}--- Fig.~\ref{fig:I1}
shows the average loss time of electrons as a function of their phase
space location in a globally stochastic magnetic field. Here the
electron losses are either through spatial transport to the vessel
wall or collisional slowing down to the bulk plasma.  It is evident
that passing electrons with $\left| \xi \right| \gtrsim 0.5$ are
rapidly lost along the open magnetic field lines, but electrons
initially located within the trapped region of momentum space can
remain confined for a far longer period of time. This is due to
magnetically trapped electrons only following a magnetic field line
for a short distance before being reflected and retracing the same
magnetic field line, thus sharply limiting their spatial
transport. The loss time for such magnetically trapped electrons is
determined by their detrapping rate. Since the collisional detrapping
rate decreases rapidly with the electron's energy, relativistic
trapped electrons can remain confined in the plasma for an extended
period, even for a fully stochastic magnetic field. This population
thus provides a remnant seed capable of surviving global MHD
instabilities.  A primary aim of this work will be to evaluate the
magnitude and decay rate of this seed across a range of plasma
conditions, thus providing guidance on what experimental conditions
must be achieved to reduce this seed to a specific level.

{\it Self-Consistent RE Evolution.}--- Our simulation model includes a
drift kinetic description of runaway electron
evolution~\cite{mcdevitt2019avalanche}, a power balance equation, and
a flux diffusion equation, all in a toroidal plasma with nested
circular flux surfaces.  Specifically, for the poloidal magnetic flux
$\psi,$
\begin{equation}
\left. \frac{\partial \psi}{\partial t} \right|_{r} =
\frac{\eta}{\mu_0} \frac{1}{\left\langle R^{-2} \right\rangle}
\frac{1}{r} \frac{\partial}{\partial r} \left[ r \left\langle R^{-2}
  \right\rangle \frac{\partial \psi}{\partial r} \right] -
\frac{\eta}{B_0 R_0} \frac{\left\langle \mathbf{j}_{RE} \cdot
  \mathbf{B} \right\rangle}{\left\langle R^{-2} \right\rangle}
, \label{eq:ME1}
\end{equation}
with $\eta$ the background plasma resistivity and
$\left\langle \cdots \right\rangle$ a flux surface average.  Ohm's
law takes the form $E_\Vert = \eta \left( j_\Vert - j_{RE}\right)$
with $j_{RE}$ the runaway current density. A conducting wall boundary
condition $\psi(r=a)=0$ is imposed. The power balance for background
plasma follows
\begin{equation}
\frac{3}{2} \frac{\partial p}{\partial t} = \frac{1}{r}
\frac{\partial}{\partial r} \left( r n \chi \frac{\partial
  T_e}{\partial r} \right) - S_{rad} \left( n_e, T_e\right) + S_{RE} +
\frac{E^2_\Vert}{\eta} . \label{eq:ODFM1}
\end{equation}
Here electrons, ions and neutrals are assumed to have the same
temperature, $p = n_e T + T \sum_j n_j$ is the total background plasma
pressure with the sum over all ion and neutral species, $\chi$ is the
heat diffusivity taken to be $\chi = 1\;\text{m}^2/\text{s}$,
$S_{rad}$ describes radiative losses, $S_{RE}$ is the energy gained by
the bulk plasma due to REs slowing down against free electrons, and
the last term describes Ohmic heating. The charge state and $S_{rad}$
are evaluated using data generated by the collisional radiative code
FlyChk~\cite{chung2005flychk} under
the assumption of steady state.

{\it Simulation set up.}---The first step of the simulation study is
the preparation of a runaway current plateau. This is obtained through an
idealized thermal quench that imposes a cooling history of $T(r,t) =
\left[ T_{init} \left( r\right) - T_{final} \left( r\right) \right]
\exp \left( -t/\Delta t_{TQ} \right) + T_{final} \left( r\right)$,
with the final temperature $T_{final} \left( r\right)
= T_{f} \left[ 1-0.7 \left( r/a\right)^2 \right]$ and $T_{f}$ the
on-axis temperature after the thermal quench. The initial temperature
and density profiles follow $T_{init} \left( r\right) = T_{0} \left[
  1-0.7 \left( r/a \right)^2 \right]^2$ and $n_{e,init} \left(
r\right) = n_{D,init} \left( r\right) = n_{D0} \left[ 1-0.9 \left( r /
  a\right)^2 \right]^{2/3}$, where $T_{0}$ and $n_{D0}$ are the
on-axis temperature and deuterium density, respectively.

For all cases considered, we will assume $T_0=3.1\;\text{keV}, T_f=10\;\text{eV}$, and
$n_{D0}=2.8\times 10^{13}\;\text{cm}^{-3}$. Once the hot tail seed has
formed, it is amplified by the avalanche mechanism leading to the
formation of a RE plateau, where the amount of RE current is
controlled by varying the amount of initial plasma current. After the
Ohmic to RE current conversion is complete, large quantities of
material are injected into the plasma at two different times. During
the first injection, a variable amount of neon is injected into the RE
beam. By injecting the neon into an existing RE beam, we have
the freedom to vary the amount of RE current and injected neon
independently.

Once the neon enters the plasma, the temperature is evaluated via the
power balance equation [Eq. (\ref{eq:ODFM1})]. A second injection
composed entirely of deuterium is made later in the simulation,
resulting in a factor of ten increase in the deuterium
density. Shortly after the deuterium injection, a fraction $f_{purge}$
of the initial neon is removed from the plasma in order to describe
the purge of high-Z material, a phenomenon observed across a range of
experiments~\cite{shiraki2018dissipation, pautasso2020generation,
  hollmann2020study, reux2022physics}.

\begin{figure}
\begin{centering}
\subfigure[]{\includegraphics[scale=0.25]{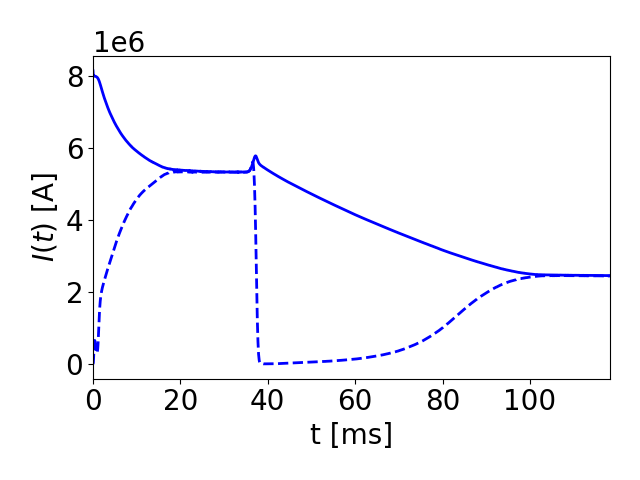}}
\subfigure[]{\includegraphics[scale=0.25]{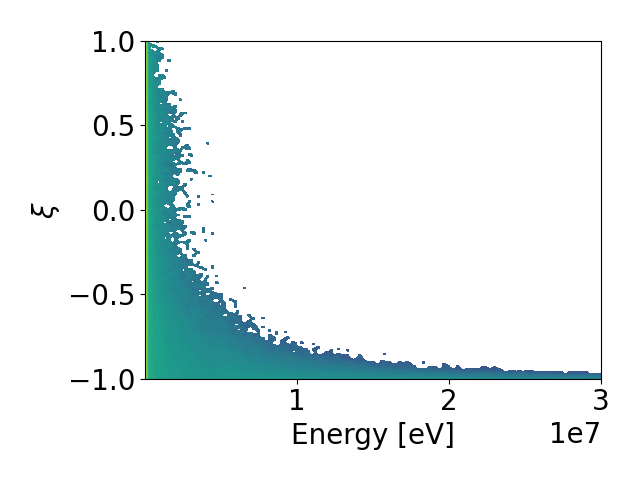}}
\subfigure[]{\includegraphics[scale=0.25]{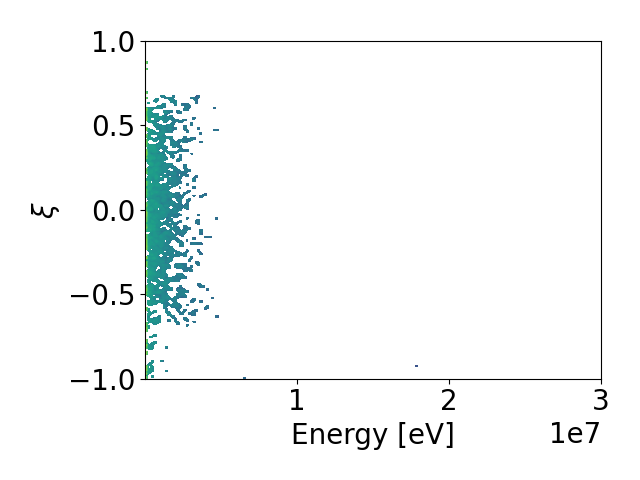}}
\subfigure[]{\includegraphics[scale=0.25]{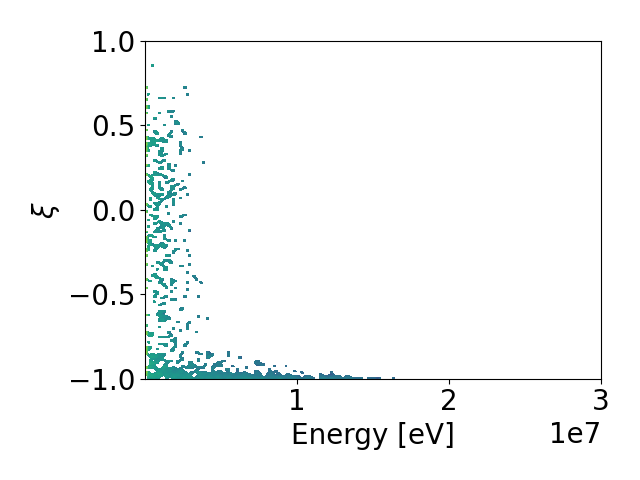}}
\par\end{centering}
\caption{(a) The evolution of plasma current versus time. The solid curve indicates the total current, whereas the dashed curved indicated the non-thermal current. Time slices
  of the momentum distribution of REs before
  [$t-t_{TQ}\approx 33\;\text{ms}$, panel (b)], during
  [$t-t_{TQ}\approx 39\;\text{ms}$, panel (c)] and after
  [$t-t_{TQ}\approx 47\;\text{ms}$, panel (c)] the MHD instability. The
  plasma was assumed to initially carry $I_p \approx 8\;\text{MA}$,
  minor radius $a = 200\;\text{cm}$, an on-axis magnetic field
  $B_0=3\;\text{T}$, $f_{purge}=0.9$, and $\Delta t_{open} \approx
  1.43\;\text{ms}$.}
\label{fig:SEMHD1kinetic}
\end{figure}

{\it Runaway Flush \& Remnant Seed.}---The RE plateau thus formed
provides the initial condition for investigating the 3D
MHD flush of runaways.  Consistent with the toroidally-averaged
formulation of Eq.~(\ref{eq:ME1}), we model the enhanced loss of
runaways in a globally stochastic magnetic field by a spatial
diffusivity of the form,
\begin{equation}
D^{kin}_{RE} \left( \gamma, \xi, r, \theta \right) = \frac{\left|
  v_\Vert \right|}{c} \Theta \left[ \left| \xi \right| - \xi_{trap}
  \left( r, \theta \right) \right] D_{RE} , \label{eq:SEMHDkinetic1a}
\end{equation}
where $\xi_{trap} \left( r, \theta \right) = 1 - B \left( r, \theta
\right)/B_{Max}$, $B \left( r, \theta \right)$ is the
magnetic field strength at the electron's current location, $B_{Max}$ is the
maximum value of the magnetic field on the flux surface,
and $D_{RE}=\left( 1/4\right) \left(1/N_t \right) \left(
a^2/\tau_{transit}\right)$.  Here, $\tau_{transit}=2\pi R_0/c$ is the
time it takes a relativistic electron to make one toroidal transit,
$N_{transit}$ is the number of toroidal transits made by a stochastic magnetic
field line while traversing the plasma, where this latter quantity is
used to parameterize the strength of the spatial transport. In order
to describe the increase in RE transport induced by the large, though
transient MHD instability, the diffusivity will be assumed to follow
$D_{RE} \left( t\right) = D^{kin}_{RE} \exp \left[ -\left(
  t-t_{open}\right)^2 / \Delta t^2_{open} \right]$, where $t_{open}$
is the time at which the RE transport is the largest, and $\Delta
t_{open}$ sets the duration of the MHD event.

An example of the impact of a global MHD instability on the phase
space distribution of REs is shown in
Fig. \ref{fig:SEMHD1kinetic}. Here, the MHD instability reaches its
peak amplitude at $t\approx35.8\;\text{ms}$, resulting in the loss of
nearly the entire RE current. From Figs. \ref{fig:SEMHD1kinetic}(b)
and \ref{fig:SEMHD1kinetic}(c), it is evident that the expulsion of
the relativistic electron population is not complete. In particular, a
significant number of trapped relativistic electrons remain confined,
providing a seed RE population after the MHD instability ceases. This
relativistic population of trapped electrons emerges due to the strong
pitch-angle scattering~\cite{Hesslow:2017} coinciding with the
presence of a high-Z material such as neon during the current
plateau~\cite{mcdevitt2018spatial, liu2021compressional}. Once the
flux surfaces reheal, the detrapping of this remnant population of
trapped electrons (typically less than one percent of the initial RE
population), allows for a sizable RE seed to robustly form [see
  Fig.~\ref{fig:SEMHD1kinetic}(d)]. This seed is subsequently amplified by
the avalanche mechanism allowing for the partial reformation of the RE
plateau.  In the following we will identify key parameters that
influence $j^{(seed)}_{RE}$ and $\psi_{10}$, and thus runaway
reconstitution.

\begin{figure}
\begin{centering}
\subfigure[]{\includegraphics[scale=0.25]{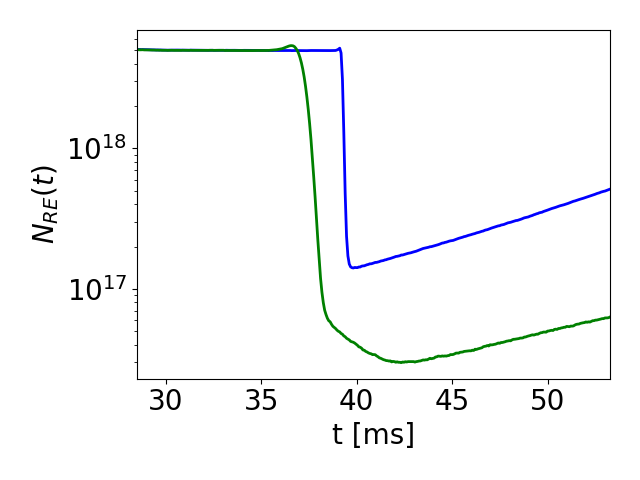}}
\subfigure[]{\includegraphics[scale=0.25]{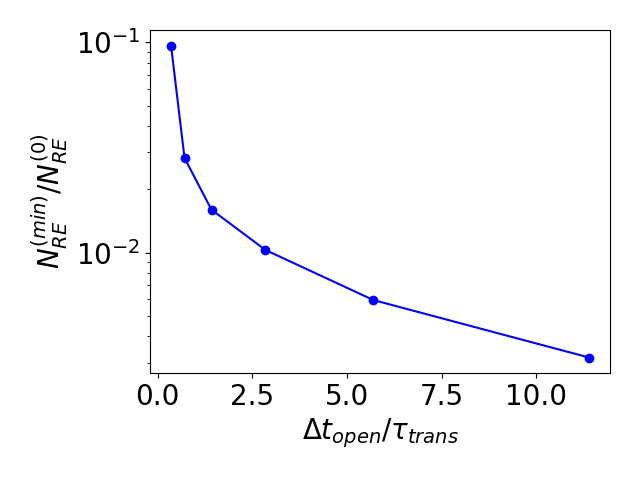}}
\par\end{centering}
\caption{Panel (a) shows the evolution of the number of energetic electrons for $\Delta t_{open} \approx 1.79\times 10^{-4}\;\text{s}$ (blue curve) and $\Delta t_{open} \approx 1.43 \times 10^{-3}\;\text{s}$ (green curve). Panel (b) shows the fraction of surviving RE electrons versus the duration of the MHD instability $\Delta t_{open}$ normalized to the transport time scale $\tau_{trans}\equiv a^2/D_{RE}$. The other parameters are $I_p \approx 8\;\text{MA}$, minor radius $a\approx 200\;\text{cm}$, $B_0=3\;\text{T}$, $n_{Ne} = 2n_{D0}$, and $f_{purge} = 0.9$.}
\label{fig:SEMHD1kineticsub0}
\end{figure}

First assessing the impact of the duration of the MHD instability
$\Delta t_{open}$ on the size of the remnant seed population,
Fig.~\ref{fig:SEMHD1kineticsub0}(a) shows the number of energetic
electrons that remain confined in the plasma for two different values
of $\Delta t_{open}$. The shortest value of $\Delta t_{open}$ is
chosen to be comparable to the transport time scale $\tau_{trans} =
a^2/D_{RE}$ of REs by the 3D magnetic field. For this case, it is
evident that the RE population drops sharply during the period of
enhanced transport, which is a result of the rapid loss of passing
REs. In contrast, considering a case where $\Delta t_{open} \gg
\tau_{trans}$, the number of REs drops sharply initially, but then
transitions to a period with a far slower decay rate. These two
different decay rates are linked to the different rates that passing
and trapped relativistic electrons are lost in a stochastic magnetic
field. In particular, while passing electrons can be directly lost via
free streaming along open magnetic field lines, trapped electrons are
not directly lost. Instead magnetically trapped electrons must first
be detrapped, before they may exit the plasma via streaming along open
magnetic field lines. For a relativistic electron population, this
detrapping rate is far slower compared to the time scale associated
with transport along open magnetic field lines, thus resulting in the
slow decay rate of the RE population evident in
Fig.~\ref{fig:SEMHD1kineticsub0}(a). This transition from a rapid
decay rate for $\Delta t_{open} \lesssim \tau_{trans}$ to a much
slower decay rate for $\Delta t_{open} \gg \tau_{trans}$ is
illustrated in Fig.~\ref{fig:SEMHD1kineticsub0}(b).

\begin{figure}
\begin{centering}
\subfigure[]{\includegraphics[scale=0.25]{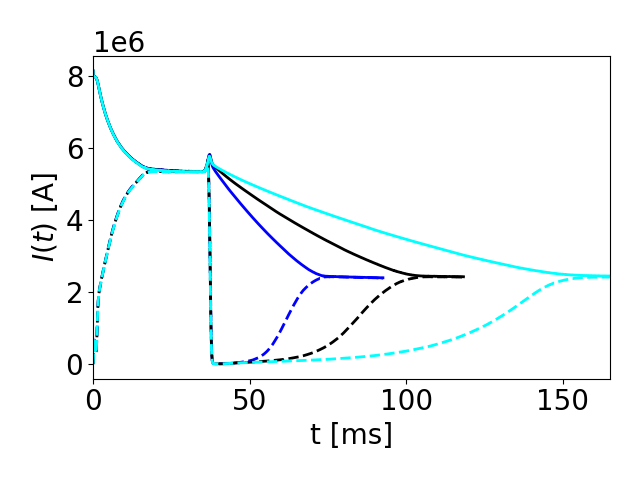}}
\subfigure[]{\includegraphics[scale=0.25]{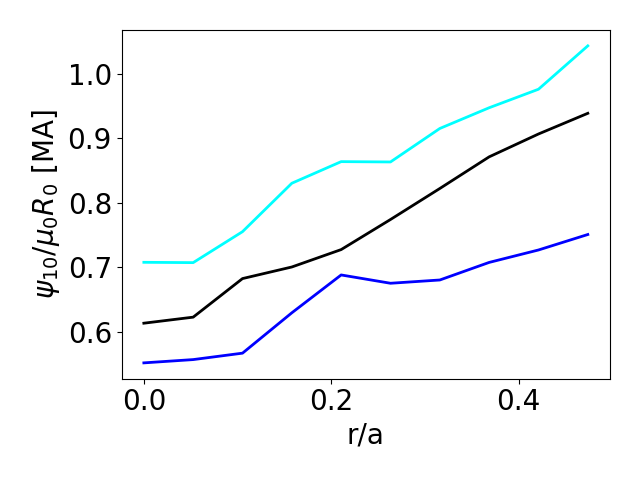}}
\par\end{centering}
\caption{Evolution of the plasma current [panel (a)] and the inferred $\psi_{10}$ for different purge fractions $f_{purge}$. The blue curves are for $f_{purge}=0.6$, the black curves are for $f_{purge}=0.9$, and the blue curves are for $f_{purge}=0.975$. The other parameters are, $I_p \approx 8\;\text{MA}$, minor radius $a\approx 200\;\text{cm}$, $B_0=3\;\text{T}$, $n_{Ne} = 2n_{D0}$, and $\Delta t_{open} \approx 1.43\;\text{ms}$.}
\label{fig:SEMHD3}
\end{figure}

{\it Runaway Reconstitution.}---We will also be interested in
identifying the impact of the amount of neon remaining in the plasma
after the deuterium injection. Three cases with different values of
$f_{purge}$ are indicated in Fig. \ref{fig:SEMHD3}. Here, it is
apparent that in all three cases a similar amount of RE current is
able to reform. The origin of this apparent insensitivity is the
result of two partially offsetting physical processes. The first is
that the amount of impurities that remain in the plasma increases the
detrapping rate of the remnant population by two distinct
mechanisms. The first is due to the higher neon content increasing the
collisional detrapping rate. A more subtle mechanism is due to the
larger inductive electric field present in the higher neon density
case, due to the plasma being radiatively pinned to a lower
temperature. This larger inductive electric field induces a Ware pinch
of the trapped electron population, which convects the electrons
inward where they are more easily detrapped. Hence, the
retention of neon in the plasma leads to a reduction of the remnant
seed population. In contrast, the presence of neon increases the
efficiency of the avalanche mechanism. This trend is shown in
Fig. \ref{fig:SEMHD3}(b), where as the quantity of neon is increased,
the value of $\psi_{10}$ decreases, implying less poloidal flux is
required to increase the runaways by an order of magnitude. For the
present example, these competing effects largely offset one another,
yielding a weaker than expected sensitivity to $f_{purge}$.

\begin{figure}
\begin{centering}
\subfigure[]{\includegraphics[scale=0.25]{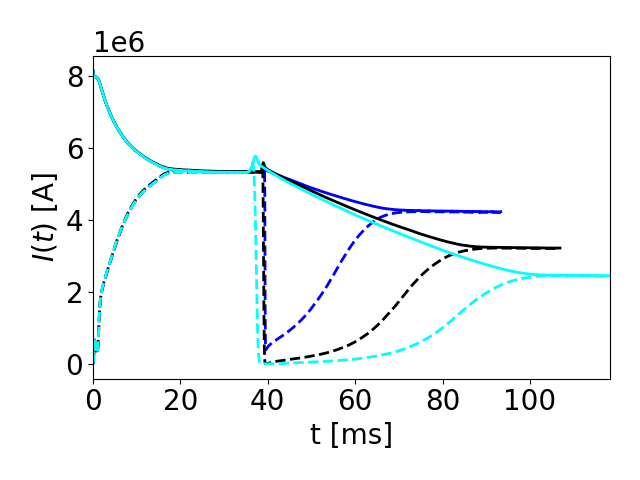}}
\subfigure[]{\includegraphics[scale=0.25]{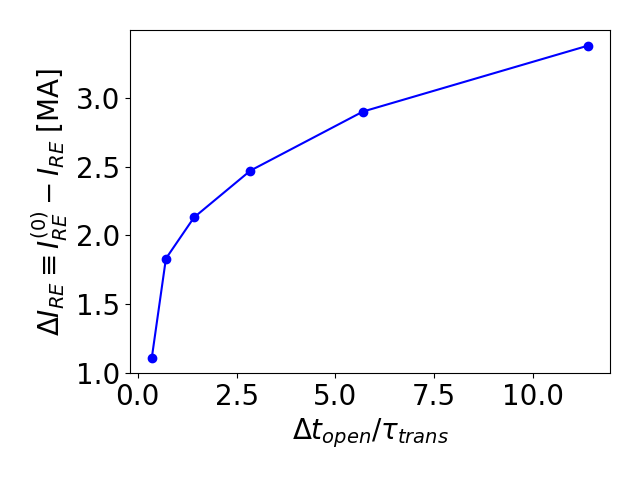}}
\par\end{centering}
\caption{Current evolution [panel (a)] and the size of the plasma current drop versus the time scale the flux surfaces remain open [panel (b)] plotted on a linear scale. The curves correspond to $\Delta t_{open} \approx 0.09\;\text{ms}$ (blue curve), $\Delta t_{open} \approx 0.36\;\text{ms}$ (black curve), and $\Delta t_{open} \approx 1.43\;\text{ms}$ (cyan curve). The initial Neon density was $n_{Ne0}=2n_{D0}=5.6\times 10^{13}\;\text{cm}^{-3}$, $a=200\;\text{cm}$, $B_0 =3\;\text{T}$ and $\tau_{trans}\equiv a^2/D_{RE}\approx 0.26 \;\text{ms}$.}
\label{fig:IMHD1}
\end{figure}

Finally, we will investigate the impact of the duration of the MHD
event on runaway reconstitution.  The amount of reformed
RE current for three different values of $\Delta t_{open}$ is shown in
Fig. \ref{fig:IMHD1}(a). When increasing $\Delta t_{open}$ from
$0.09\;\text{ms}$ to $1.43\;\text{ms}$, the amount of reformed RE
current is observed to decrease from $3.9\;\text{MA}$ to
$2.0\;\text{MA}$. Hence, while the time scale that the flux surfaces
remain open is increased by a factor of sixteen, the amount of
reformed RE current is only reduced by roughly a factor of two. This
relative insensitivity is due to the slow decay rate of the trapped
remnant energetic electron population, along with the exponential
dependence of the avalanche amplification mechanism on the amount of
poloidal flux consumed. The resulting drop in RE current as a function
of $\Delta t_{open}$ is shown in Fig. \ref{fig:IMHD1}(b), where it is
apparent that the current drop becomes a weak function of $\Delta
t_{open}$ once $\Delta t_{open} \gg \tau_{trans}$.

{\it Discussion.} Runaway reconstitution after a 3D MHD flush is
surprisingly robust because of a new seeding mechanism
via ``trapped runaways.''  While the size of this seed varies with the
amount of neon initially injected into the plasma, the purge fraction,
and the time ($\Delta t_{open}$) that flux surfaces remain open, its
decay is ultimately set by the detrapping from a combination of
pitch-angle scattering and the Ware pinch. For the range of parameters
considered here, the magnitude of this trapped remnant seed ranged
from roughly 1\% to 0.1\% of the initial RE population. For most cases
of interest $\psi_{10}/\mu_0 R_0 \lesssim 1\;\text{MA}$,
so runaway reconstitution comes with a 2-3 MA plasma current drop.
Impeding runaway reconstitution depends on (1) enhancing the
detrapping rate and (2) prolonging $\Delta t_{open}.$ Increasing the
post-purge $n_D$ is a straightforward approach for (1), although it is
constrained by the assimilation of injected deuterium into the plasma
and the density window~\cite{paz2021novel, BatteyREM2022} identified
for triggering the 3D MHD flush.
Experimental studies reveal significant variation in $\Delta
t_{open}$~\cite{paz2021novel}, but the precise control knob remains to
be understood.  A more reliable alternative is a passive runaway
coil~\cite{boozer2011two, weisberg2021passive, tinguely2021modeling}
that could hold the flux surfaces open for a far longer time
period.



\begin{acknowledgments}

We thank the U.S. Department of Energy Office of Fusion Energy
Sciences and Office of Advanced Scientific Computing Research
for support under the Tokamak Disruption Simulation (TDS) Scientific
Discovery through Advanced Computing (SciDAC) project, and the Base
Theory Program, both at Los Alamos National Laboratory (LANL) under
contract No. 89233218CNA000001. This research used
resources of the National Energy Research Scientific Computing Center
(NERSC), a U.S. Department of Energy Office of Science User Facility
operated under Contract No. DE-AC02-05CH11231 and the Los Alamos
National Laboratory Institutional Computing Program, which is
supported by the U.S. Department of Energy National Nuclear Security
Administration under Contract No. 89233218CNA000001.

\end{acknowledgments}

\newpage

\end{document}